\documentclass[12pt]{article}
\usepackage{longtable}
\usepackage{graphicx}
\usepackage{amsmath}
\usepackage{float}
\usepackage{booktabs}
\usepackage{geometry}
 \usepackage{amsmath}
\usepackage{tikz}
\usetikzlibrary{positioning,arrows.meta}
\usepackage{xcolor}
\usepackage{float}
 \usepackage{multirow} 
 \usepackage{tikz}
\usetikzlibrary{positioning,arrows.meta}
 \usepackage{graphicx}
\usepackage{float}
 
\usepackage{natbib}
\usepackage{tikz}
\usetikzlibrary{positioning,arrows.meta}
\usepackage{amsmath}
\usepackage{booktabs}
\usepackage{geometry}
\usepackage{graphicx}
\usepackage{graphicx}
\usepackage{float}
\usepackage{setspace}
\usepackage{tikz}
\tikzset{main/.style={circle,draw,minimum size=1cm}}
\tikzset{box/.style={draw,rounded corners,thick,align=center,minimum width=4.3cm,minimum height=1.2cm}}

\usepackage{longtable}

\usepackage{tikz}

\usetikzlibrary{
positioning,
arrows.meta,
shapes.geometric
}
 
\usepackage{authblk}
\usetikzlibrary{
positioning,
arrows.meta,
shapes.geometric,
shapes.misc
}
 \usetikzlibrary{shapes.misc}
\usepackage{amssymb}
\usepackage{array}
\usepackage{tikz}
\usepackage{longtable}
\usepackage{booktabs}
\usepackage{float}      
\usepackage{caption}    
\usepackage{subcaption} 
\usepackage{graphicx}
\title{
AI-Assisted Causal Inference and Mediation Analyses of Environmental and Psychosocial Determinants of Subjective Cognitive Difficulties in the All of Us Research Program
}

\author[1]{Cong Cao}
\author[1]{Shuangge Ma\thanks{Corresponding author: Shuangge Ma, Department of Biostatistics, Yale School of Public Health, Yale University, New Haven, CT 06520, USA. Email: \texttt{shuangge.ma@yale.edu}}}

\affil[1]{Department of Biostatistics, Yale School of Public Health, Yale University, New Haven, CT 06520, USA}

\date{} 

\begin{document}

\maketitle
\begin{abstract}
Short-term environmental exposures have been linked to cognitive and behavioral outcomes, although many reported associations may reflect broader geographic and contextual differences. Using longitudinal data from the All of Us Research Program (2018--2024), we linked daily weather and air-pollution exposures to repeated attention-related and subjective cognitive outcomes. Associations were evaluated using pooled, fixed-effects, lagged, and event-study analyses. Additional machine-learning analyses were conducted to explore potential heterogeneity and latent psychosocial structure. Replication analyses were performed using the 2024 Behavioral Risk Factor Surveillance System (BRFSS). Several environmental exposure measures showed small associations with cognitive outcomes in pooled analyses, but most attenuated substantially after accounting for within-location temporal variation. Mediation, sensitivity, and machine-learning analyses yielded similar conclusions. In contrast, mental-health burden, loneliness, and social functioning were consistently associated with subjective cognitive difficulty and exhibited substantially larger effect sizes than environmental exposures. Similar patterns were observed in BRFSS. Exploratory AI-assisted analyses yielded findings broadly consistent with the primary longitudinal analyses. These findings suggest that short-term environmental perturbations may have limited associations with cognitive outcomes after accounting for within-location variation, whereas psychosocial factors appear to be more consistently associated with subjective cognitive burden.

\end{abstract}

\vspace{0.5em}

\noindent

 
  \section*{Introduction}

Environmental conditions such as temperature extremes, air pollution, and severe weather events have been linked to cognitive performance, attention, productivity, and mental health in a wide range of observational studies \citep{currie2009air,neidell2010pollution,zhang2018impact,greenstone2015environment,hsiang2013quantifying,burke2015global}. Proposed mechanisms include physiological stress responses, sleep disruption, mood changes, inflammation, and alterations in daily behavior. These findings have contributed to growing interest in the cognitive and behavioral consequences of environmental change.

Despite this literature, interpretation of short-term environmental associations remains challenging. Environmental exposures are strongly patterned by geography, season, infrastructure, socioeconomic conditions, and population health characteristics. Regions that experience different weather conditions often differ simultaneously in healthcare access, environmental burden, social vulnerability, and baseline health status \citep{haines2019imperative,romanello20232023,watts20212020}. Consequently, associations identified in pooled observational analyses may reflect persistent contextual differences between locations rather than acute environmental effects occurring within the same location over time.

Repeated longitudinal data provide an opportunity to address this challenge more directly. By leveraging temporal variation within geographic areas, longitudinal designs can help distinguish short-term environmental fluctuations from stable regional characteristics that may confound cross-sectional comparisons \citep{hernan2016using}. Such approaches are particularly relevant when estimated effects are modest and potentially sensitive to model specification.

These issues may be especially important for subjective cognitive and attention-related outcomes. Self-reported concentration difficulties and cognitive complaints are influenced not only by cognitive functioning itself but also by emotional well-being, fatigue, stress, social isolation, and broader psychosocial context \citep{tec2023weather2vec}. Subjective cognitive symptoms are increasingly recognized as clinically meaningful indicators of cognitive vulnerability and have been associated with elevated risk of subsequent cognitive decline in population-based studies. At the same time, psychosocial factors such as loneliness, mental-health burden, and impaired social functioning are among the strongest known correlates of subjective cognitive symptoms. Evaluating these factors alongside environmental exposures may therefore provide important context for interpreting observed environmental-health associations.

In this study, we used longitudinal data from the All of Us Research Program to examine whether short-term environmental conditions are associated with subjective cognitive difficulties. Repeated observations enabled a series of increasingly restrictive identification strategies designed to distinguish associations driven by persistent geographic and contextual differences from those arising from within-location environmental variation over time. By comparing estimates across pooled, fixed-effects, lagged, and event-based analyses, we evaluated the robustness of environmental associations under progressively stronger causal identification assumptions. We also examined a broad range of psychosocial, behavioral, physiological, and environmental factors associated with subjective cognitive symptoms. Together, these analyses provide a more comprehensive assessment of factors related to subjective cognitive difficulties in a large and geographically diverse U.S. population.

\section*{Methods}

\begin{figure}[htbp]
\centering
\includegraphics[
width=0.55\textwidth
]{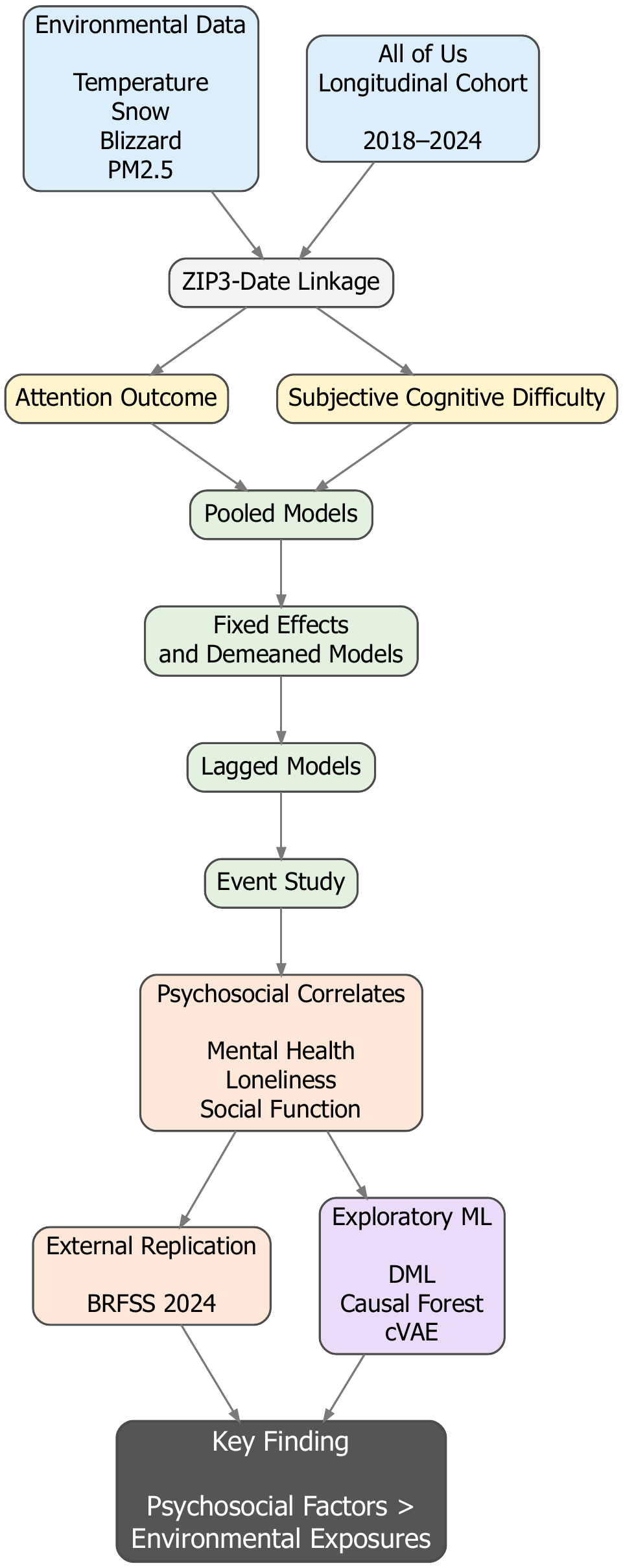}
\caption{Study workflow and analytical framework.}
\label{fig:workflow}
\end{figure}

Figure~\ref{fig:workflow} summarizes the study workflow and analytical framework. We examined whether short-term environmental perturbations were associated with repeated attention-related and subjective cognitive outcomes and explored psychological, behavioral, physiological, and environmental factors that could help interpret observed associations.
The study focused on short-term environmental variation rather than long-term climate differences across regions. Daily environmental exposure records were linked to repeated participant observations using assessment dates and ZIP3 geographic information. The longitudinal structure of the data allowed us to examine whether changes in environmental conditions were associated with changes in attention-related and subjective cognitive outcomes within the same geographic areas over time.

\subsection*{Data sources and cohort construction}
 
The analytic dataset was created by linking repeated participant-level observations from the All of Us Research Program collected between 2018 and 2024 with daily environmental exposure records using ZIP3 geographic information and observation dates. The All of Us Research Program operates under central Institutional Review Board (IRB) oversight, and all participants provided informed consent for research participation. The present study used de-identified data accessed through the All of Us Researcher Workbench in accordance with program policies and data-use requirements. Participant-level data included subjective cognitive and attention-related outcomes, demographic characteristics, behavioral measures, psychological assessments, socioeconomic variables, and selected laboratory biomarkers. Environmental data included daily temperature, snowfall, precipitation, blizzard exposure, and PM$_{2.5}$ measures derived from weather and air-pollution datasets.

The environmental exposure dataset contained approximately 2.18 million ZIP3-date observations. The primary analyses included up to 427,127 participant-day observations for attention-related outcomes and up to 369,508 participants for harmonized subjective cognitive difficulty outcomes. Supplementary analyses involving alternative cognitive outcome definitions included up to 454,661 participants. Environmental exposures were assigned according to participant ZIP3 location and observation date. Sample size varied across analyses because data availability differed across environmental exposures, behavioral measures, psychological assessments, and laboratory biomarkers.

 Rather than focusing solely on absolute temperature levels, we evaluated multiple measures of short-term environmental variation, including temperature anomalies, temperature shocks, lagged exposure measures, cumulative exposure windows, and extreme-weather indicators. Exposure measures included daily minimum temperature (TMIN), maximum temperature (TMAX), snowfall (SNOW), and precipitation (PRCP), together with relative measures constructed from local temperature anomalies, temperature deviations, and standardized snow-related variables.

To examine short-term environmental perturbations, we additionally evaluated temperature shocks, snow shocks, day-to-day exposure differences, rolling exposure windows, and lagged exposure measures. These variables captured departures from typical local conditions as well as abrupt changes occurring over short periods of time. Additional analyses considered blizzard exposure, extreme cold days, extreme heat days, and snow-related threshold indicators. PM$_{2.5}$ was linked using the same spatial and temporal resolution and evaluated as an additional environmental exposure. Detailed exposure definitions and construction procedures are provided in Supplementary Appendix A. Geographic variation in environmental exposures across study regions is summarized in Supplementary figure~\ref{fig:weather_map}.

The primary objective was to distinguish associations driven by differences between locations from associations arising from environmental variation within the same locations over time. This framework can be viewed as a quasi-experimental causal inference strategy that progressively reduces confounding from stable geographic and contextual factors. To address this question, we compared a series of increasingly restrictive model specifications, including pooled models, fixed-effects models, doubly demeaned analyses, lagged exposure models, and event-based analyses. Extreme cold days were defined as days with minimum temperature (TMIN) below the 5th percentile of the pooled daily temperature distribution across study locations between 2018 and 2024. Extreme heat days were defined analogously using the 95th percentile of the corresponding temperature distribution. Both measures were coded as binary indicators. These analyses evaluated whether short-term weather events were associated with changes in attention-related and subjective cognitive outcomes. Psychological, behavioral, physiological, and environmental variables were evaluated as potential correlates and candidate mediators.

\subsection*{Identification strategy}

To distinguish between cross-sectional geographic differences and within-location environmental variation, we adopted a fixed-effects framework that exploits temporal fluctuations in environmental conditions within the same geographic area over time. Let \(Y_{zt}\) denote the cognitive outcome observed in ZIP3 region \(z\) at time \(t\), and let \(E_{zt}\) denote the corresponding environmental exposure. The primary fixed-effects specification was

\begin{equation}
Y_{zt}
=
\beta E_{zt}
+
\alpha_z
+
\gamma_t
+
\varepsilon_{zt},
\end{equation}

where \(\alpha_z\) represents ZIP3-specific fixed effects and \(\gamma_t\) represents calendar-time effects.

This specification removes time-invariant differences across locations, including stable socioeconomic, infrastructural, climatic, and demographic characteristics. Under the assumption that short-term environmental fluctuations are not systematically correlated with unmeasured time-varying determinants of cognition after conditioning on location and time effects, \(\beta\) can be interpreted as the association arising from within-location environmental variation over time.

To evaluate temporal dynamics surrounding short-term environmental events, we estimated event-study models of the form

\begin{equation}
Y_{zt}
=
\sum_{k\neq -1}
\beta_k D_{z,t+k}
+
\alpha_z
+
\gamma_t
+
\varepsilon_{zt},
\end{equation}

where \(D_{z,t+k}\) denotes indicators for days occurring \(k\) periods before or after an environmental event, and the period immediately preceding the event (\(k=-1\)) serves as the reference category.

Coefficients for \(k<0\) were used to assess pre-event trends, whereas coefficients for \(k>0\) quantified post-event deviations from baseline. The absence of systematic pre-event trends would be consistent with the assumption that exposed and unexposed periods followed similar trajectories prior to environmental perturbations.

\subsection*{Independent replication analyses using BRFSS}

To evaluate whether psychosocial correlates identified in the All of Us cohort were reproducible in an independent population-based sample, supplementary analyses were conducted using data from the 2024 Behavioral Risk Factor Surveillance System (BRFSS). The primary replication outcome was subjective cognitive decline (SCD), defined using the BRFSS item assessing worsening confusion or memory loss during the previous 12 months (CIMEMLO1). Additional outcomes included decision-making difficulty (DECIDE), cognitive-related worry (CDWORRY), social limitation attributable to cognitive symptoms (CDSOCIA1), and household limitation attributable to cognitive symptoms (CDHOUS1). Psychosocial variables included poor mental-health days during the previous month (MENTHLTH), loneliness (SDLONELY), life satisfaction (LSATISFY), and history of depression (ADDEPEV3). Logistic regression models adjusted for age, sex, race, education, and household income were used to evaluate associations between psychosocial variables and cognitive outcomes. Exploratory mediation analyses were conducted using loneliness as exposure, poor mental health days as a mediator, and subjective cognitive decline as the result. Because BRFSS is cross-sectional, mediation findings were interpreted with caution and not considered evidence of causal pathways.

The heterogeneity of the potential effect was evaluated in age, smoking exposure, alcohol use, elderly status, and socioeconomic subgroups using interaction models and subgroup analyses. Supplementary analyses were conducted using double machine learning (DML), causal forests, and related semiparametric approaches to evaluate potential nonlinear relationships and high-dimensional confounding. Exploratory latent-variable analyses using conditional variational autoencoders (cVAEs) were also conducted to examine a lower-dimensional psychosocial and behavioral structure related to subjective cognitive difficulty. These analyses were treated as robustness analyses and were used to assess whether conclusions differed from those obtained using the primary longitudinal regression models. Effect estimates are reported with 95\% confidence intervals and two-sided \(p\)-values. Detailed replication results are presented in Table~\ref{tab:brfss_replication}
\section*{Results}

\subsection*{Environmental associations attenuate under within-location analyses}

Across both attention and subjective cognitive difficulty outcomes, environmental associations were generally small and attenuated under identification strategies emphasizing within-location temporal variation.

For attention outcomes, pooled logistic regression models identified a modest positive association between temperature-shock exposure and attention impairment ($\beta=0.0315$, $p=0.032$). However, estimated coefficients were substantially attenuated in ZIP3 fixed-effects, doubly demeaned ZIP-by-month, ZIP-level aggregated, and elderly-restricted fixed-effects models, where estimates ranged from 0.001 to 0.011 and were not statistically significant. Lagged exposure analyses likewise yielded small and imprecise estimates ($\beta=0.0498$, $p=0.270$). Supplementary robustness analyses incorporating demographic adjustment and nonlinear exposure-response terms produced findings similar to those of the primary model.

\begin{figure}[p]
\centering

\begin{subfigure}[t]{0.69\textwidth}
\centering
\includegraphics[width=\textwidth]{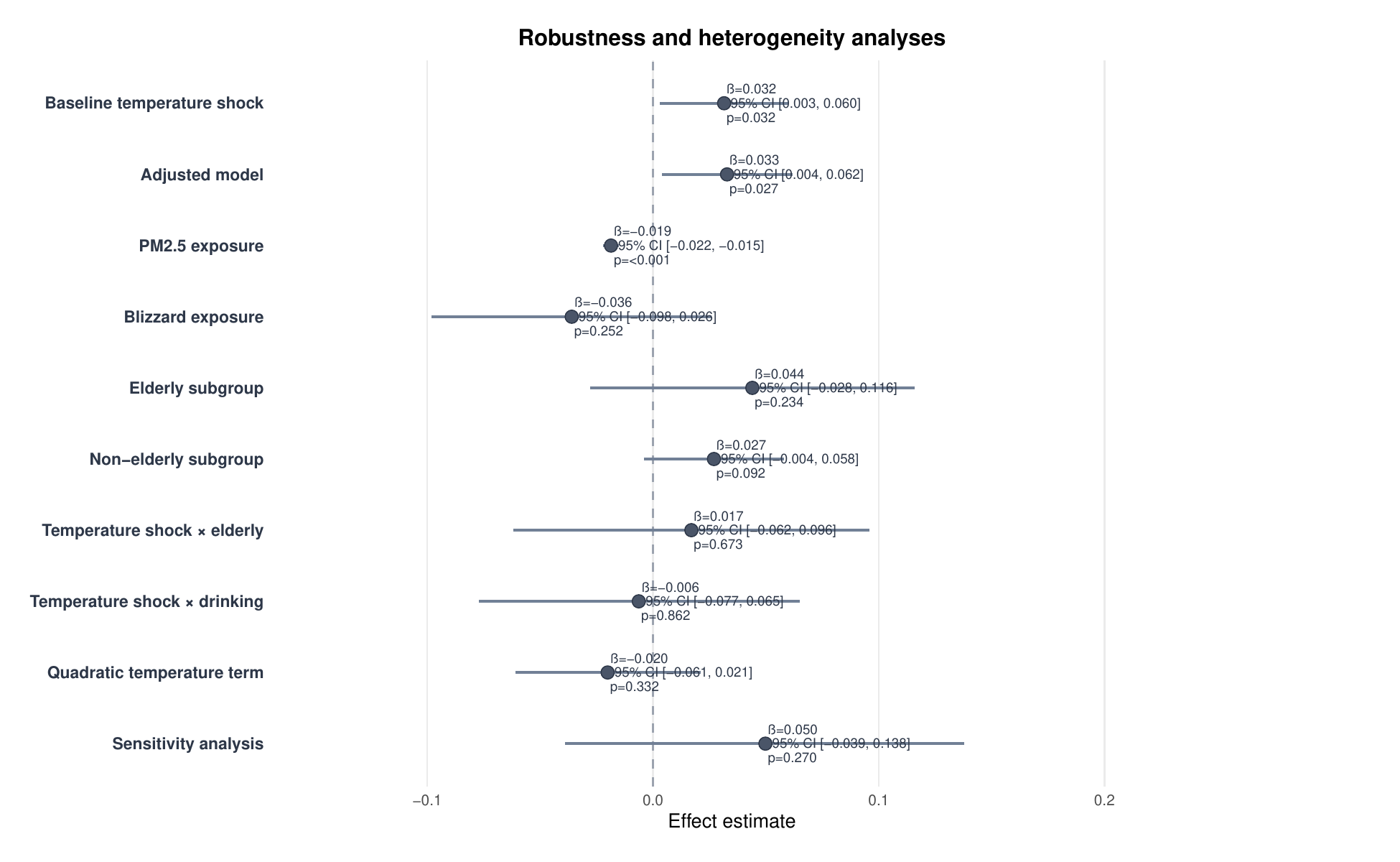}
\caption{Robustness analyses.}
\end{subfigure}
\hfill
\begin{subfigure}[t]{0.49\textwidth}
\centering
\includegraphics[width=\textwidth]{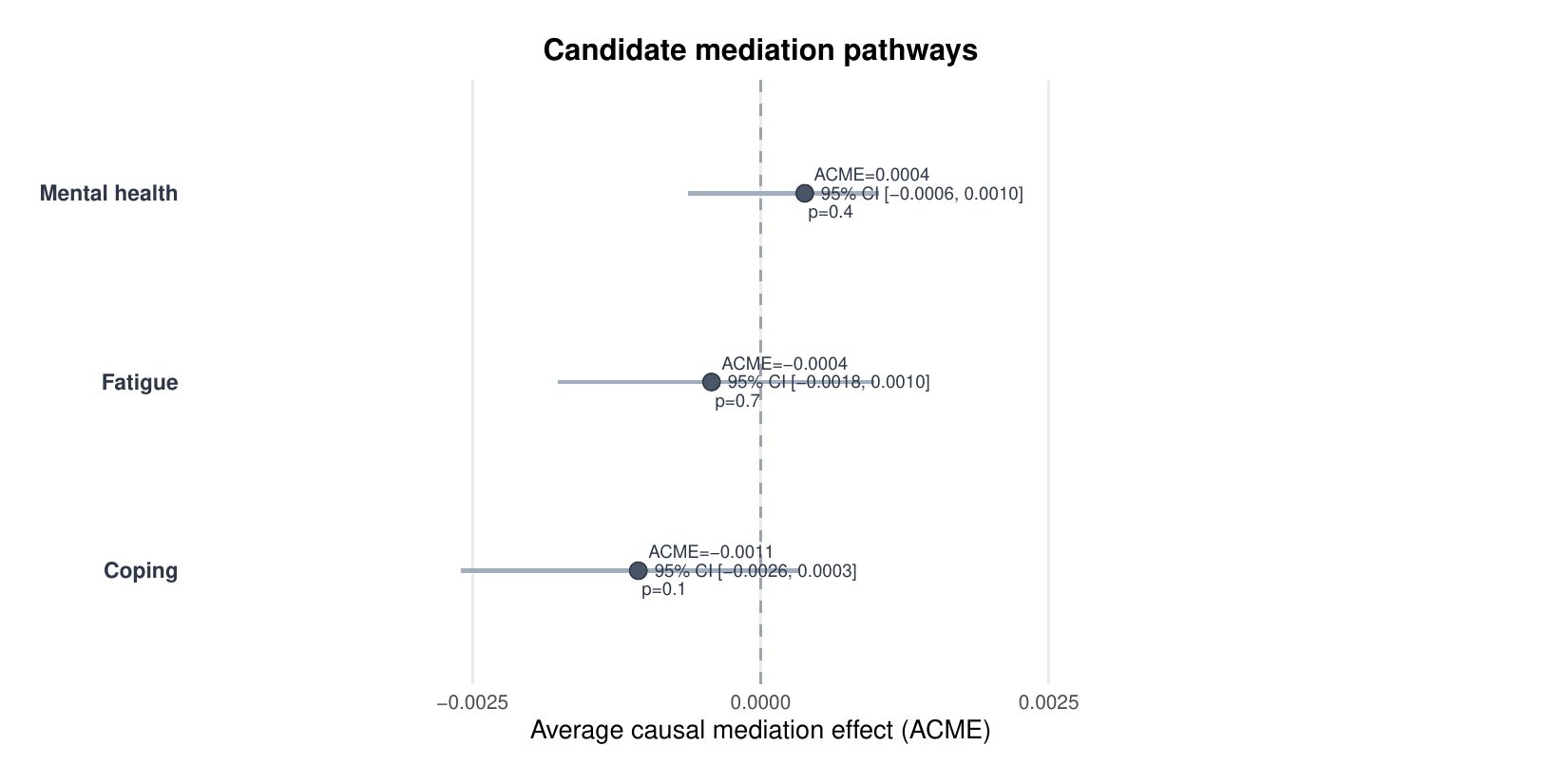}
\caption{Mediation analyses.}
\end{subfigure}

\vspace{0.5em}

\begin{subfigure}{0.50\textwidth}
\centering
\includegraphics[width=\textwidth]{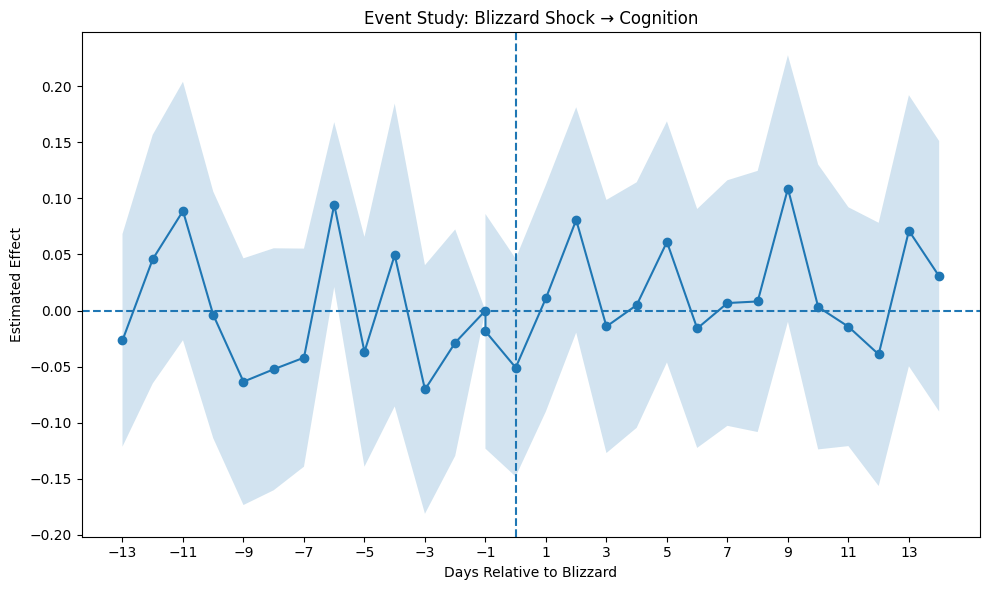}
\caption{Event-study estimates.}
\end{subfigure}

\caption{
\textbf{Supplementary Figure S2. Environmental effects were generally weak across robustness, temporal, and mediation analyses.}
Panel A summarizes robustness analyses across alternative model specifications. Panel B presents event-study estimates surrounding short-term environmental exposure periods. Panel C summarizes mediation analyses of candidate pathways linking environmental exposures with attention outcomes. Across analyses, estimated effects were generally small and confidence intervals frequently included the null.
}

\label{fig:environmental_effects_summary}

\end{figure}

Associations between environmental exposures and subjective cognitive difficulty were similarly weak. Pooled, fixed-effects, ZIP-level aggregated, and doubly demeaned models all produced estimates close to zero and were not statistically significant. Event-study analyses showed no clear pre-exposure trends, immediate post-exposure increases, or sustained post-exposure deviations. Lagged exposure models and analyses restricted to older participants yielded similar results. Supplementary Figure~\ref{fig:environmental_effects_summary} summarizes robustness, event-study, and mediation analyses.

Exploratory mediation analyses yielded uniformly small indirect effects across candidate psychological, physiological, behavioral, and environmental pathways. PM$_{2.5}$ exposure was negatively associated with attention outcomes ($\beta=-0.0186$, $p<0.001$), whereas blizzard exposure was not associated with attention impairment ($\beta=-0.0360$, $p=0.252$) or subjective cognitive difficulty ($\beta=-0.0003$, $p=0.652$). Interaction and subgroup analyses provided little evidence of systematic effect heterogeneity across age, drinking status, smoking exposure, or elderly subgroups. Estimated interaction coefficients were small and confidence intervals consistently included the null.

Collectively, these findings suggest that associations between short-term environmental perturbations and cognitive outcomes were modest in magnitude and weakened substantially under within-location identification strategies.

\begin{figure}[H]
\centering
\includegraphics[width=\textwidth]{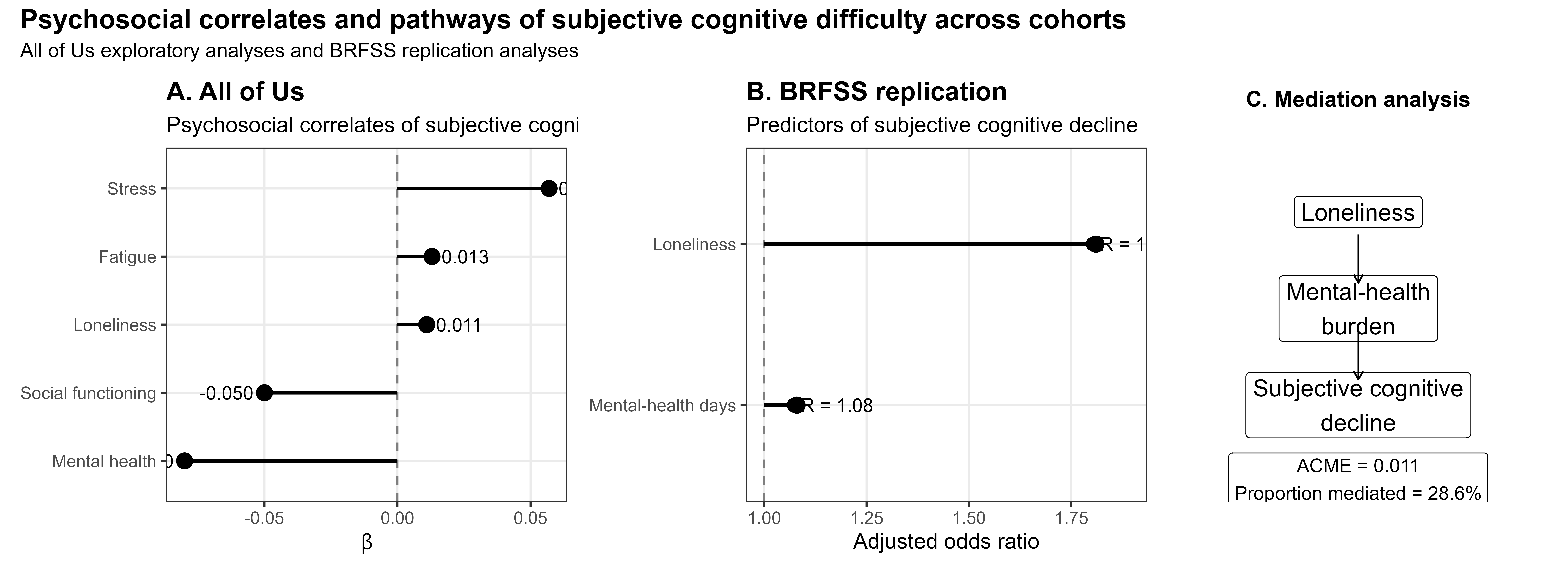}
\caption{
Psychological and psychosocial correlates of subjective cognitive difficulty across the All of Us and BRFSS cohorts. Psychosocial measures demonstrated substantially larger effect sizes than environmental exposure measures and showed consistent associations across independent datasets.}
\label{fig}
\end{figure}
 \subsection*{Psychosocial correlates dominate environmental exposures}

Psychological and psychosocial measures showed substantially stronger associations with subjective cognitive difficulty than environmental exposure variables (Figure~\ref{fig}). Mental-health burden demonstrated the largest association with subjective cognitive difficulty ($\beta=-0.114$, $p<0.001$), followed by social functioning ($\beta=-0.102$, $p<0.001$), literacy-related measures ($\beta=-0.082$, $p<0.001$), and loneliness ($\beta=0.053$, $p<0.001$). In contrast, cumulative blizzard exposure remained close to zero ($\beta=-0.0003$, $p=0.652$). Loneliness was also associated with greater cognitive difficulty ($\beta=0.011$, $P<0.001$). Higher stress levels were associated with greater cognitive difficulty ($\beta=0.057$, $P<0.001$), whereas fatigue showed a smaller but statistically significant association ($\beta=0.013$, $P<0.001$). Overall, psychological and psychosocial variables consistently exhibited substantially larger associations with subjective cognitive difficulty than short-term environmental exposures.

Blizzard exposure was not associated with mental-health measures in either residualized or cumulative exposure models.
Indirect effects involving mental health, PM$_{2.5}$, smoking behavior, and alcohol use were generally small across specifications. Predicted mental-health values from first-stage models were likewise not strongly associated with subjective cognitive difficulty. Similar findings were observed across psychological, physiological, and environmental mediation analyses. Overall, the mediation analyses provided limited evidence that short-term environmental exposures influenced subjective cognitive difficulty through indirect psychological, behavioral, physiological, or environmental pathways.

\begin{figure}[htbp]
\centering
\includegraphics[
width=1\textwidth
]{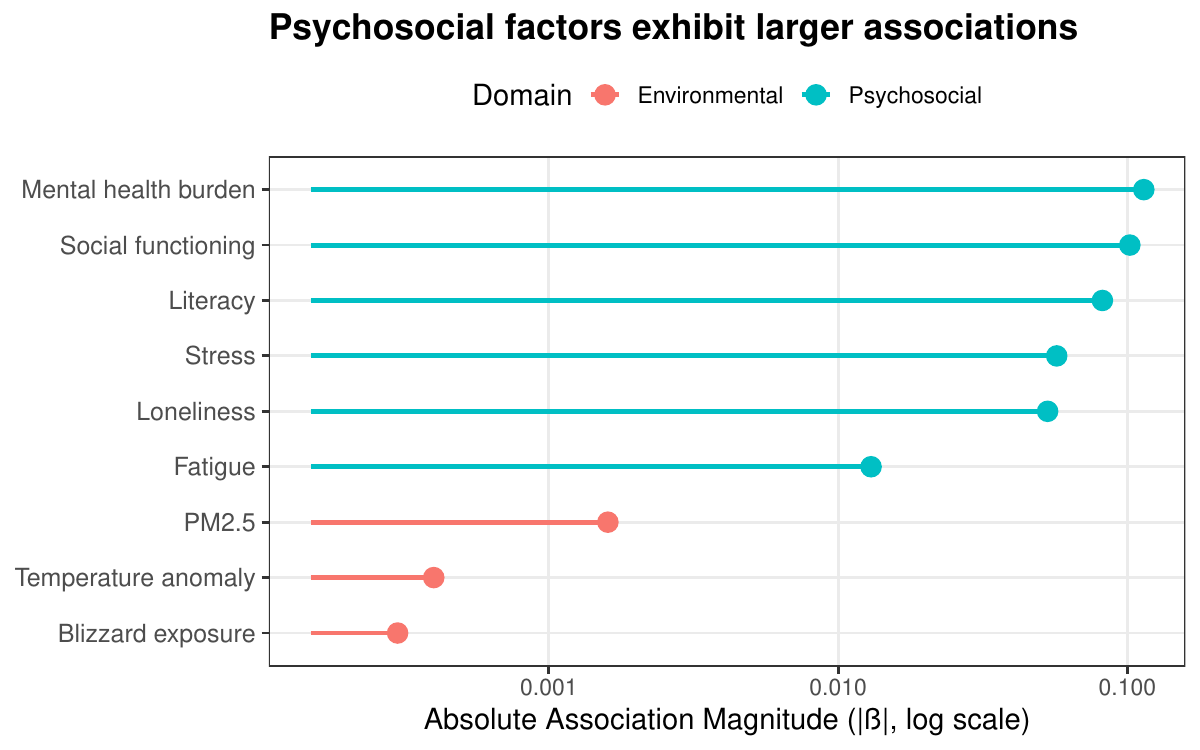}

\caption{
Relative magnitude of psychosocial and environmental associations with subjective cognitive difficulty.
Psychosocial measures, including mental-health burden, social functioning, loneliness, stress, and literacy-related measures, exhibited substantially larger associations with subjective cognitive difficulty than short-term environmental exposures. The logarithmic scale highlights the large separation in effect-size magnitude between psychosocial and environmental domains.
}

\label{fig:effect_size_ranking}
\end{figure}

Psychological variables, including loneliness, social health, and mental-health measures, were summarized using a conditional variational autoencoder (cVAE). Temperature shock was treated as the exposure and subjective cognitive difficulty as the outcome. Exploratory latent-representation analyses produced findings broadly consistent with the primary analyses. Higher latent psychological burden was generally associated with greater subjective cognitive difficulty. Additional analyses involving heterogeneous treatment effects, PM$_{2.5}$ interactions, transformer-based semantic embeddings, and extended latent-representation models are reported in the Supplementary Appendix. To evaluate whether nonlinear relationships, latent psychosocial structure, or heterogeneous treatment effects altered the primary conclusions, we conducted a series of exploratory AI-assisted analyses, including double machine learning, causal forests, conditional variational autoencoders, and embedding-based approaches. Findings were broadly consistent with the primary analyses and are reported in Supplementary Appendix D. Psychological and psychosocial measures showed substantially stronger associations with subjective cognitive difficulty than environmental exposure variables (Figure~\ref{fig:effect_size_ranking}).
 
\subsection*{Independent replication analyses using BRFSS}
To evaluate whether psychosocial correlates identified in the All of Us cohort generalized to an independent population-based sample, we conducted supplementary analyses using data from the 2024 Behavioral Risk Factor Surveillance System (BRFSS). BRFSS is a nationwide telephone-based health survey coordinated by the U.S. Centers for Disease Control and Prevention (CDC) that collects information on health conditions, health-related behaviors, and social determinants of health among U.S. adults. Among respondents reporting subjective cognitive symptoms, loneliness and poor mental-health days were associated with cognitive-related worry, social limitations, and household limitations. Higher loneliness scores were associated with greater odds of cognitive-related worry (OR = 1.30, 95\% CI: 1.24--1.37), social limitation (OR = 1.22, 95\% CI: 1.15--1.29), and household limitation (OR = 1.21, 95\% CI: 1.15--1.28). Similarly, increasing numbers of poor mental-health days were associated with higher odds of cognitive-related worry (OR = 1.02 per day, 95\% CI: 1.02--1.03), social limitation (OR = 1.04 per day, 95\% CI: 1.04--1.05), and household limitation (OR = 1.04 per day, 95\% CI: 1.03--1.04).

Both direct and indirect pathways were statistically significant. The average causal mediation effect (ACME) was 0.011 (95\% CI: 0.010--0.012), whereas the average direct effect (ADE) was 0.027 (95\% CI: 0.026--0.028). The estimated total effect was 0.038 (95\% CI: 0.038--0.039). Approximately 28.6\% of the association between loneliness and subjective cognitive decline was mediated through poor mental-health days.

\begin{table}[htbp]
\centering
\caption{Replication and functional-consequence analyses in BRFSS}
\label{tab:brfss_replication}
\small
\begin{tabular}{lcc}
\toprule
Outcome &
Loneliness OR (95\% CI) &
Mental-health days OR (95\% CI) \\
\midrule

\multicolumn{3}{l}{\textbf{Replication outcome}} \\

Subjective cognitive decline
&
1.81 (1.76--1.86)
&
1.08 (1.08--1.08)
\\

\midrule

\multicolumn{3}{l}{\textbf{Functional consequences among respondents with cognitive symptoms}} \\

Cognitive worry
&
1.30 (1.24--1.37)
&
1.02 (1.02--1.03)
\\

Social limitation
&
1.22 (1.15--1.29)
&
1.04 (1.04--1.05)
\\

Household limitation
&
1.21 (1.15--1.28)
&
1.04 (1.03--1.04)
\\

\bottomrule
\end{tabular}

\vspace{0.4em}

\footnotesize{
All associations were statistically significant ($p<0.001$). Odds ratios were estimated from logistic regression models using 2024 BRFSS data. The first panel evaluates replication of psychosocial associations with subjective cognitive decline. The second panel evaluates functional consequences among respondents reporting cognitive symptoms.
}
\end{table}

\section*{Discussion and conclusion}

Using longitudinal data from the All of Us Research Program, we evaluated associations between short-term environmental perturbations and repeated attention-related and subjective cognitive outcomes across multiple complementary analytic frameworks. Environmental associations were generally modest in pooled analyses and attenuated substantially under specifications emphasizing within-location temporal variation. Importantly, attenuation occurred consistently as analyses moved toward designs with stronger causal identification assumptions. In contrast, psychological and psychosocial factors demonstrated larger and more consistent associations with subjective cognitive difficulty. Similar psychosocial patterns were observed in an independent BRFSS replication cohort, supporting the robustness of these findings across datasets and study designs.

For attention-related outcomes, pooled analyses identified small associations between temperature perturbation measures and attention impairment. However, these associations weakened considerably in fixed-effects, lagged, and related within-location analyses. A similar pattern was observed for subjective cognitive difficulty, where estimated environmental effects remained close to zero across pooled, fixed-effects, ZIP-level aggregated, event-study, and doubly demeaned specifications. Because these analyses were conducted in a large longitudinal cohort with substantial within-location variation, the attenuation of effect estimates is unlikely to be explained solely by reduced statistical power. Instead, the findings suggest that some associations observed in pooled analyses may partly reflect persistent geographic, seasonal, or contextual differences across locations rather than acute environmental effects occurring within locations over time. The consistency of results across multiple identification strategies strengthens this interpretation.

In contrast, psychosocial measures demonstrated substantially stronger and more reproducible associations with subjective cognitive difficulty. Mental-health burden, loneliness, social functioning, and related psychosocial indicators consistently emerged among the strongest correlates in the All of Us cohort. These findings were independently replicated in BRFSS, where loneliness and poor mental-health burden were strongly associated with subjective cognitive decline, cognitive-related worry, and functional limitations attributable to cognitive symptoms. The replication of these patterns across two large population-based datasets suggests that psychosocial vulnerability may play a more important role in subjective cognitive burden than short-term environmental fluctuations.

The mediation analyses provided little evidence that short-term environmental perturbations influenced cognitive outcomes through the psychological, behavioral, physiological, or environmental pathways examined. Although several candidate mediators were associated with cognitive outcomes, their associations with environmental perturbations were generally weak, resulting in small indirect effects. By contrast, exploratory BRFSS analyses suggested that mental-health burden partially mediated the association between loneliness and subjective cognitive decline, indicating a potential psychosocial pathway linking social isolation and subjective cognitive symptoms. Because these analyses were observational and mediation estimates were not causally identified, they should be interpreted as descriptive rather than mechanistic evidence.

Several limitations should be considered. First, lagged analyses were conducted in a smaller subset of observations and therefore cannot completely exclude modest delayed environmental effects. Second, the outcomes were based primarily on self-reported cognitive and attention-related measures rather than objective neuropsychological assessments. Third, environmental exposures were assigned using ZIP3-level geographic linkage and may not fully capture individual mobility, indoor environmental conditions, or adaptive behavioral responses. Residual confounding and exposure misclassification also remain possible, particularly for psychosocial and behavioral variables collected across partially overlapping survey instruments and assessment periods.

The study also has several important strengths. These include the use of repeated longitudinal observations, temporally aligned environmental exposure records, independent replication analyses in a second national cohort, and multiple complementary identification strategies designed to distinguish within-location temporal variation from broader contextual differences across locations. The convergence of findings across these approaches increases confidence in the overall conclusions.

From a public-health and health-informatics perspective, these findings highlight the importance of distinguishing environmental signals from contextual and psychosocial influences when evaluating cognitive-health outcomes in large observational datasets. Across both cohorts, psychosocial vulnerability was more consistently associated with subjective cognitive burden than with short-term environmental perturbations. Interventions targeting loneliness, psychological distress, and impaired social functioning may therefore have greater potential to reduce subjective cognitive burden than approaches focused solely on short-term environmental fluctuations. Future studies incorporating higher-resolution exposure assessment, wearable and digital cognitive monitoring, and more detailed behavioral data may help clarify whether specific environmental perturbations contribute to short-term cognitive-state changes in vulnerable populations.
\section*{Acknowledgements}
We gratefully acknowledge All of Us participants for their contributions, without whom this research would not have been possible. We also thank the National Institutes of Health’s All of Us Research Program for making available the participant data examined in this study.

\section*{Funding}

This research received no specific grant from any funding agency in the public, commercial, or not-for-profit sectors.

\section*{Ethics Approval and Consent to Participate}

The All of Us Research Program is overseen by a central Institutional Review Board (IRB), the Copernicus Group IRB (Protocol Number: 20191049). The present study used de-identified participant data accessed through the All of Us Researcher Workbench under the All of Us data access and use policies. According to the All of Us Research Program data passport model, registered researchers do not require separate IRB approval from the All of Us Research Program for individual research projects conducted within the Researcher Workbench. All participants provided informed consent at enrollment in the All of Us Research Program.

\section*{Data and Code Availability}
Data used in this study were obtained through the National Institutes of Health All of Us Research Program Researcher Workbench using the Controlled Tier dataset (CDR version 8). BRFSS data are publicly available from the U.S. Centers for Disease Control and Prevention. 
Code used to generate the analyses and figures is available at:
https://github.com/congca/latent-mediation-multimorbidity

\section*{Competing Interests}

The authors declare no competing interests.
\section*{Author Contributions}
C.C. performed the analyses, prepared the figures, and drafted the manuscript. C.C. and S.M. developed the study design, interpreted the results, and revised the manuscript. All authors reviewed and approved the final manuscript.
\bibliographystyle{plainnat}
{\small
\bibliography{sample}

@article{currie2009air,
  title={Air pollution and infant health: Lessons from New Jersey},
  author={Currie, Janet and Neidell, Matthew},
  journal={Journal of Health Economics},
  volume={28},
  number={3},
  pages={688--698},
  year={2009},
  publisher={Elsevier}
}

@article{neidell2010pollution,
  title={Pollution, health, and academic performance},
  author={Neidell, Matthew},
  journal={Journal of Environmental Economics and Management},
  volume={60},
  number={2},
  pages={119--137},
  year={2010},
  publisher={Elsevier}
}

@article{greenstone2015environment,
  title={Environmental policy, air pollution, and infant mortality in India},
  author={Greenstone, Michael and Hanna, Rema},
  journal={American Economic Review},
  volume={105},
  number={2},
  pages={364--420},
  year={2015},
  publisher={American Economic Association}
}

@article{hsiang2013quantifying,
  title={Quantifying the influence of climate on human conflict},
  author={Hsiang, Solomon M and Burke, Marshall and Miguel, Edward},
  journal={Science},
  volume={341},
  number={6151},
  pages={1235367},
  year={2013},
  publisher={American Association for the Advancement of Science}
}

@article{burke2015global,
  title={Global non-linear effect of temperature on economic production},
  author={Burke, Marshall and Hsiang, Solomon and Miguel, Edward},
  journal={Nature},
  volume={527},
  number={7577},
  pages={235--239},
  year={2015},
  publisher={Nature Publishing Group}
}

@article{zhang2018impact,
  title={The impact of exposure to air pollution on cognitive performance},
  author={Zhang, Xin and Chen, Xi and Zhang, Xiaobo},
  journal={Proceedings of the National Academy of Sciences},
  volume={115},
  number={37},
  pages={9193--9197},
  year={2018},
  publisher={National Academy of Sciences}
}

@article{romanello20232023,
  title={The 2023 report of the Lancet Countdown on health and climate change: the imperative for a health-centred response in a world facing irreversible harms},
  author={Romanello, Marina and Di Napoli, Claudia and Green, Carole and Kennard, Harry and Lampard, Pete and Scamman, Daniel and Walawender, Maria and Ali, Zakari and Ameli, Nadia and Ayeb-Karlsson, Sonja and others},
  journal={The Lancet},
  volume={402},
  number={10419},
  pages={2346--2394},
  year={2023},
  publisher={Elsevier}
}

@article{haines2019imperative,
  title={The imperative for climate action to protect health},
  author={Haines, Andy and Ebi, Kristie},
  journal={New England journal of medicine},
  volume={380},
  number={3},
  pages={263--273},
  year={2019},
  publisher={Mass Medical Soc}
}

@article{watts20212020,
  title={The 2020 report of The Lancet Countdown on health and climate change: responding to converging crises},
  author={Watts, Nick and Amann, Markus and Arnell, Nigel and Ayeb-Karlsson, Sonja and Beagley, Jessica and Belesova, Kristine and Boykoff, Maxwell and Byass, Peter and Cai, Wenjia and Campbell-Lendrum, Diarmid and others},
  journal={The lancet},
  volume={397},
  number={10269},
  pages={129--170},
  year={2021},
  publisher={Elsevier}
}

@article{hernan2016using,
  title={Using big data to emulate a target trial when a randomized trial is not available},
  author={Hern{\'a}n, Miguel A and Robins, James M},
  journal={American journal of epidemiology},
  volume={183},
  number={8},
  pages={758--764},
  year={2016},
  publisher={Oxford University Press}
}

@inproceedings{tec2023weather2vec,
  title={Weather2vec: Representation learning for causal inference with non-local confounding in air pollution and climate studies},
  author={Tec, Mauricio and Scott, James G and Zigler, Corwin M},
  booktitle={Proceedings of the AAAI Conference on Artificial Intelligence},
  volume={37},
  number={12},
  pages={14504--14513},
  year={2023}
}
}

\clearpage

\section*{Supplementary Appendix A. Data Sources and Variable Construction}
\begin{figure}[H]
\centering
\includegraphics[width=\textwidth]{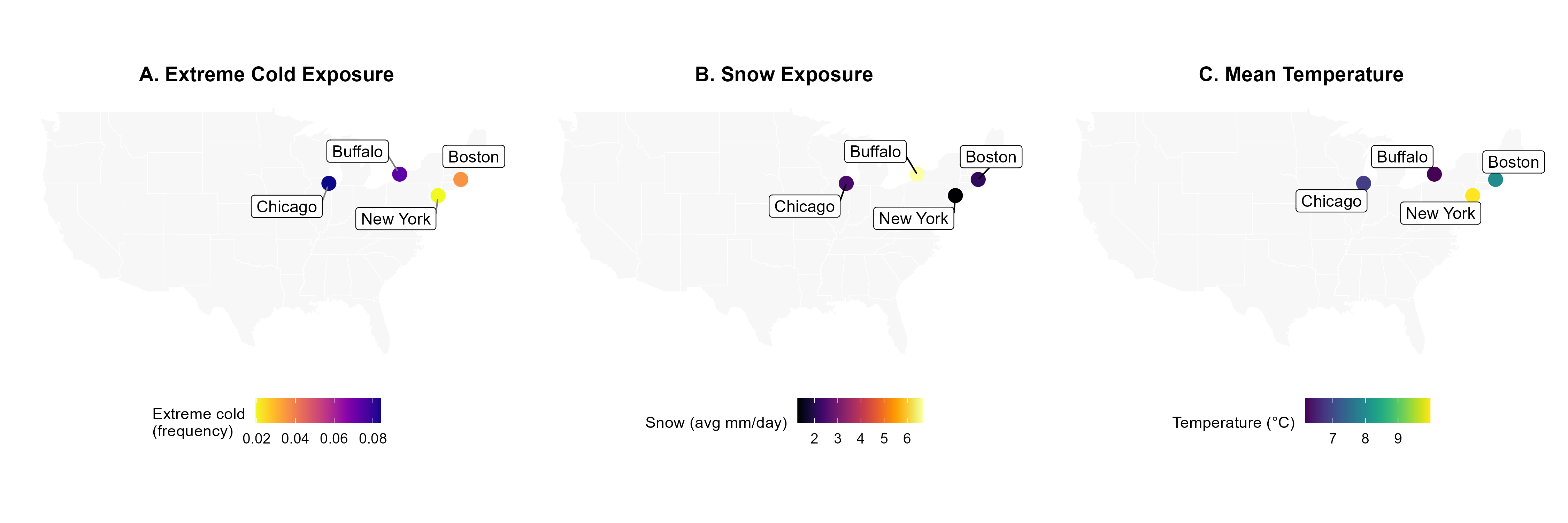}

\caption{
Geographic variation in environmental exposure characteristics across major study regions between 2018 and 2024. Panels summarize regional variation in temperature conditions, snowfall burden, and extreme cold exposure frequency.
}

\label{fig:weather_map}
\end{figure}
\subsection*{A1 Cohort construction}
\begin{figure}[p]
\centering
\includegraphics[width=\textwidth]{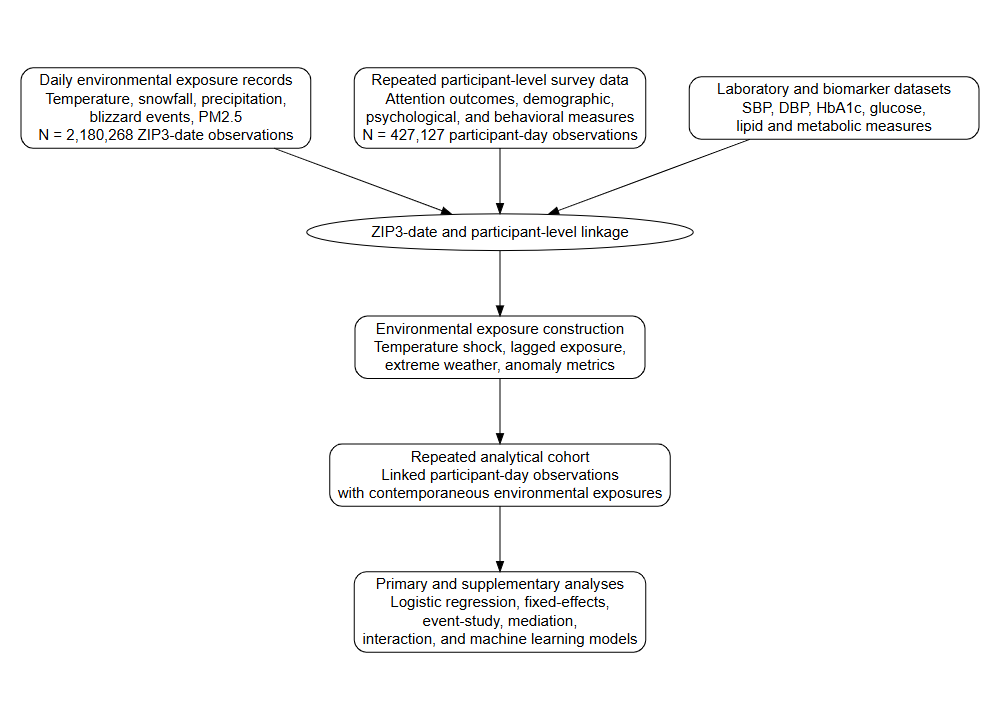}
\caption{
Overview of cohort construction and dataset linkage. Environmental, demographic, behavioral, psychological, attentional, and laboratory data were linked using participant identifiers, ZIP3 location, and observation date. Final analytical sample sizes varied across analyses because data availability differed across outcome and pathway domains.
}
\label{fig:supp_cohort_flowchart}

\end{figure}
 
\clearpage

\subsection*{A2 Variable definitions}

\scriptsize

\begin{longtable}{p{0.15\textwidth} p{0.15\textwidth} p{0.30\textwidth} p{0.30\textwidth}}

\caption{Variable definitions, source constructs, and transformations}
\label{tab:supp_variables}
\\

\toprule

\textbf{Domain}
&
\textbf{Variables}
&
\textbf{Source / Construction}
&
\textbf{Transformations}
\\

\midrule
\endfirsthead

\toprule

\textbf{Domain}
&
\textbf{Variables}
&
\textbf{Source / Construction}
&
\textbf{Transformations}
\\

\midrule
\endhead

Outcomes
&
attention (binary), attention$\_score_daily$
&
Self-reported concentration difficulty, focus-related symptoms, and repeated attention measures
&
Binary harmonization and daily aggregation
\\

&
cognition, cognition\_latent
&
Harmonized subjective cognitive difficulty measures and latent cognition constructs
&
Raw, residualized, fixed-effects, ZIP-aggregated, demeaned, and latent-variable variants
\\

\midrule

Absolute environmental exposure
&
TMIN, TMAX
&
Daily minimum and maximum temperature
&
Raw and standardized variants
\\

&
SNOW, PRCP
&
Daily snowfall and precipitation
&
Raw and lagged variants
\\

\midrule


Relative environmental anomaly
&
temp\_anomaly
&
Temperature deviation from local ZIP-specific baseline
&
ZIP-specific normalization
\\

&
snow\_z, TMIN\_z
&
Standardized snowfall and temperature measures
&
ZIP-level \(z\)-score normalization
\\

\midrule


Environmental perturbation
&
temp\_shock
&
Absolute day-to-day temperature change
&
Differenced, lagged, adjusted, and demeaned variants
\\

&
snow\_shock
&
Absolute day-to-day snowfall change
&
Standardized and lagged variants
\\

&
temp\_diff
&
Short-term temperature difference across adjacent days
&
Exposure-difference construction
\\

&
lagged exposure
&
Historical exposure measures
&
lag1--lag5 exposure propagation
\\

&
7-day exposure windows
&
Cumulative short-term environmental burden
&
Rolling aggregation across prior 7 days
\\

\midrule


Extreme-weather indicators
&
blizzard
&
Severe winter-weather event
&
Binary threshold-based indicator
\\

&
extreme\_cold
&
Daily minimum temperature below the pooled 5th percentile of the study-wide TMIN distribution (2018–2024)
&
Binary daily indicator 
\\

&
extreme\_heat
&
Daily maximum temperature above the pooled 95th percentile of the study-wide temperature distribution
&
Binary indicator
\\

&
snow\_day
&
Any measurable snowfall event
&
Binary event indicator
\\

\midrule


Environmental co-exposure
&
PM$_{2.5}$
&
Daily fine-particulate-matter exposure
&
Raw and adjusted variants
\\

\midrule
Demographic and socioeconomic variables
&
age
&
Survey-derived demographic variable
&
Continuous variable
\\

&
sex/gender, race/ethnicity
&
Survey-derived demographic responses
&
Categorical encoding
\\

&
income, education
&
Annual household income and highest educational attainment
&
Ordinal grouping and categorical harmonization
\\

\midrule
Psychological measures
&
mental\_health
&
General mental-health measures
&
Numeric and adjusted variants
\\

&
loneliness, social
&
Companionship, social isolation, and social-function measures
&
Adjusted and harmonized variants
\\

&
literacy
&
Medical-form confidence and related items
&
Numeric harmonization
\\

\midrule
Behavioral measures
&
smoking\_freq, smoking\_100
&
Smoking frequency and lifetime smoking exposure
&
Numeric and binary encodings
\\

&
alcohol
&
Alcohol-use frequency during the prior year
&
Frequency harmonization
\\

\midrule 
Physiological and laboratory measures
&
bmi
&
Clinical measurements
&
Continuous variable
\\

&
sbp, dbp
&
Blood-pressure measurements
&
Continuous variables
\\

&
hba1c, glucose
&
Metabolic biomarkers
&
Continuous variables
\\

&
creatinine
&
Renal biomarker
&
Continuous variable
\\

&
cholesterol, ldl, hdl, triglycerides
&
Lipid biomarkers
&
Continuous variables
\\

\midrule
Latent pathway constructs
&
fatigue, stress
&
Multi-item symptom and psychological-distress measures
&
Latent-variable construction
\\

&
sleep, coping
&
Sleep-related and coping-related survey items
&
Latent-variable construction
\\

\midrule

\end{longtable}

\vspace{0.5em}

\footnotesize{
Notes: Environmental exposures were assembled at the ZIP3-date level and linked to participant-level survey, behavioral, psychological, attentional, cognitive, and laboratory data using observation date and geographic assignment. Analytical sample sizes varied across models because several pathway and laboratory variables were available only in partially overlapping subsets. Outcome-specific sample sizes are reported in the corresponding regression tables.
}

\begin{figure}[htbp]
\centering
\includegraphics[width=\textwidth]{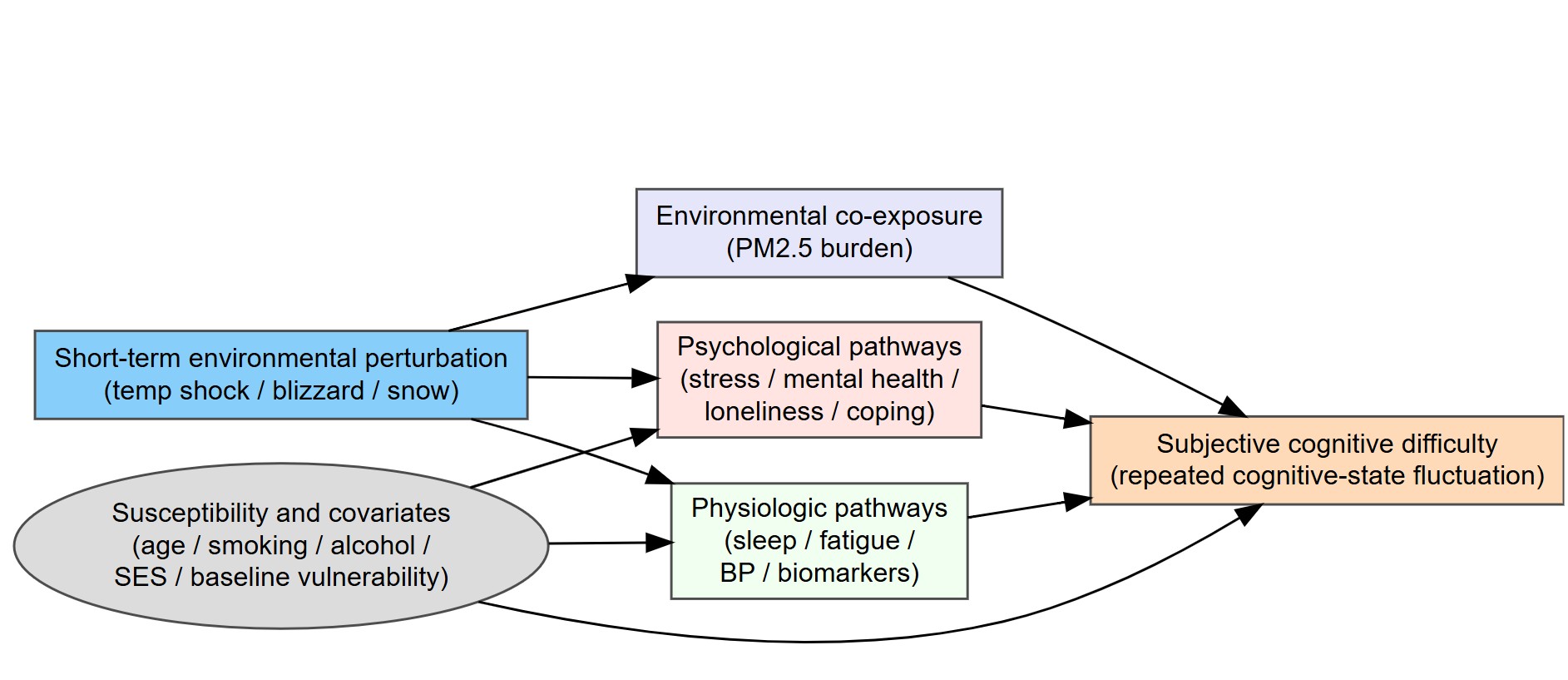}

\caption{
  Framework of environmental perturbation and repeated cognitive-attentional outcomes. Potential correlates and pathway domains included psychological, behavioral, physiological, and environmental factors.
}

\label{fig:conceptual_framework}
\end{figure}

\begin{figure}[htbp]
\centering
\includegraphics[width=\textwidth]{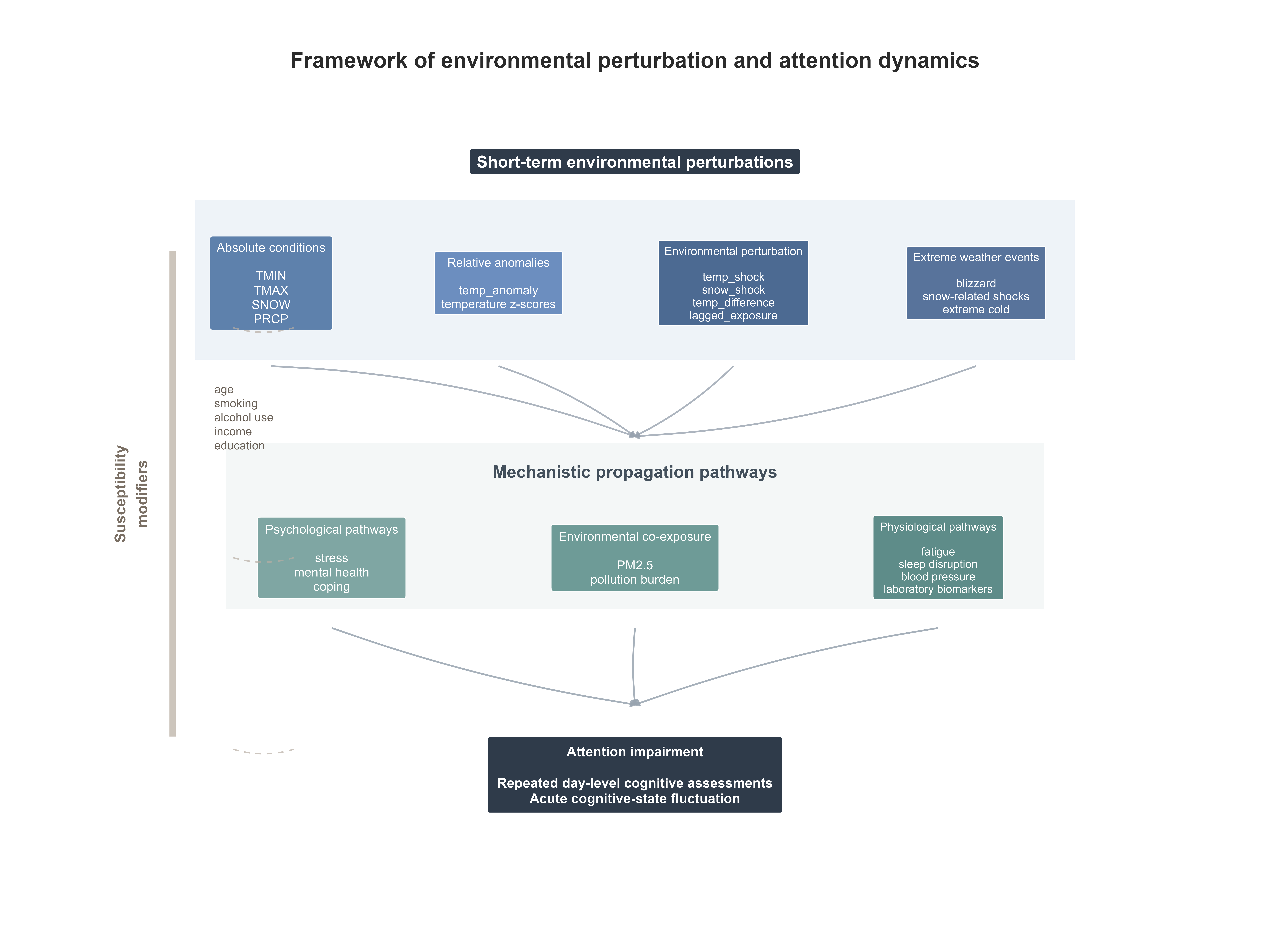}

\caption{
 Conceptual framework of environmental perturbation and repeated cognitive-attentional outcomes.
}

\label{fig:conceptual_framework}
\end{figure}

\begin{figure}[htbp]
\centering
\includegraphics[width=\textwidth]{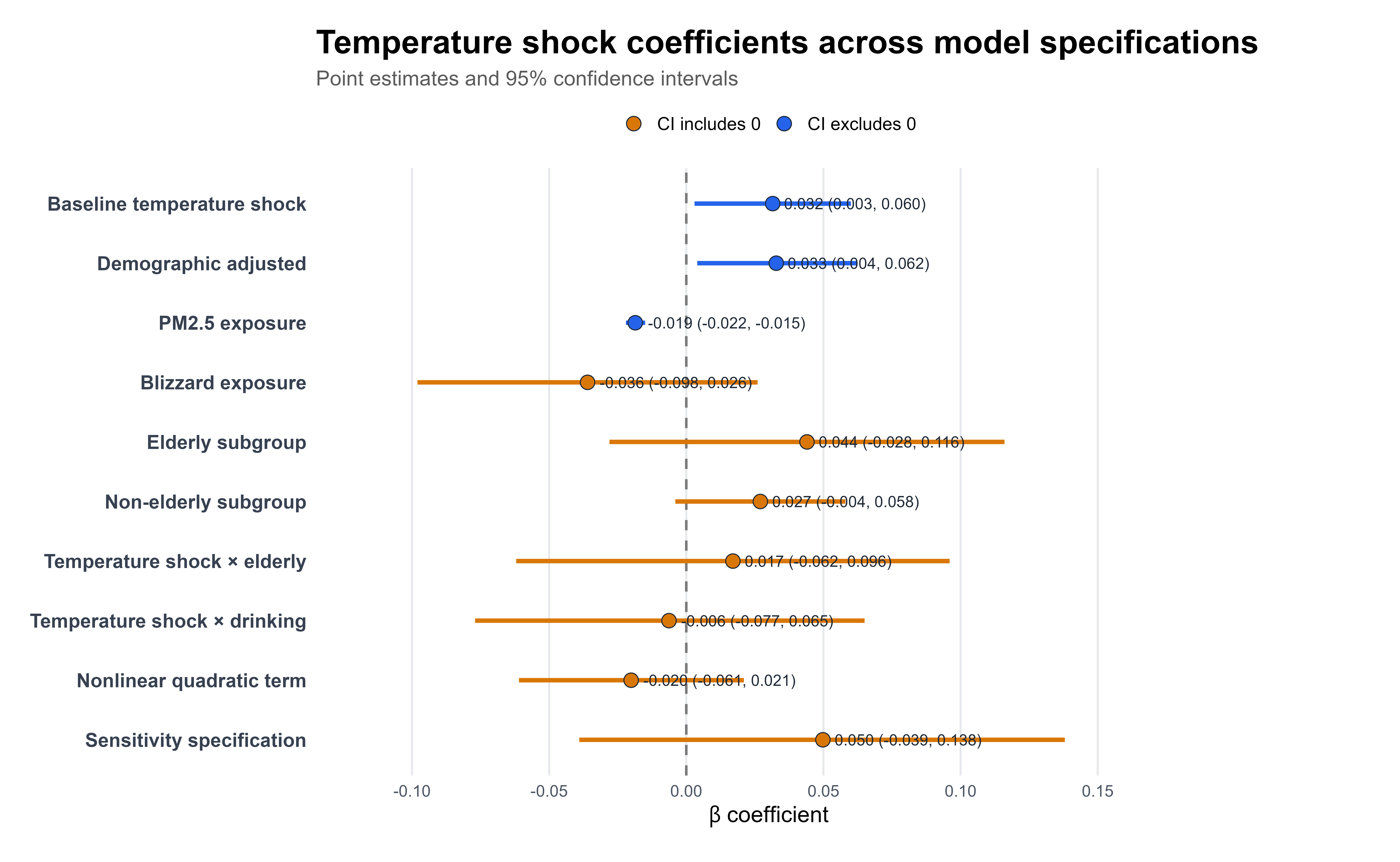}

\caption{
 Robustness results
}

\label{fig:conceptual_framework}
\end{figure}
\clearpage
\section*{Supplementary Appendix B Primary association and robustness analyses}

\begin{table}[htbp]
\centering
\caption{Primary and robustness analyses of temperature exposure and subjective cognitive difficulty}
\label{tab:cognition_all_models}
\scriptsize

\begin{tabular}{lccccc}
\toprule

Statistic
&
7-day temperature
&
Temperature shock
&
Fixed-effects
&
ZIP-aggregated
&
Two-way demeaned
\\

\midrule

Exposure variable
&
$temp\_7day\_true$
&
$temp\_shock\_r$
&
$temp\_7day$
&
$temp\_7day\_zip$
&
$temp\_7day\_zm$
\\

Coefficient
&
$9.406\times10^{-5}$
&
$-3.85\times10^{-5}$
&
$8.603\times10^{-5}$
&
$7.322\times10^{-6}$
&
$4.098\times10^{-5}$
\\

95\% CI
&
(-0.000,\;0.000)
&
($-1.07\times10^{-4}$,\;$3.02\times10^{-5}$)
&
(-0.0003,\;0.0004)
&
(-0.000,\;0.000)
&
(-0.000,\;0.000)
\\

$p$-value
&
0.503
&
0.272
&
0.617
&
0.966
&
0.813
\\

\midrule

Intercept
&
0.1197
&
$-5.13\times10^{-18}$
&
--
&
--
&
--
\\

Std. Error
&
$<0.001$
&
$3.50\times10^{-5}$
&
--
&
--
&
--
\\

t statistic
&
0.670
&
-1.099
&
--
&
--
&
--
\\

\midrule

Observations
&
315,476
&
369,508
&
367,922
&
367,922
&
367,922
\\

Entities
&
--
&
--
&
787
&
--
&
--
\\

Time periods
&
--
&
--
&
1,440
&
--
&
--
\\

\midrule

$R^2$
&
0.000
&
0.000
&
$6.806\times10^{-7}$
&
0.000
&
0.000
\\

Adjusted $R^2$
&
--
&
--
&
--
&
-0.000
&
-0.000
\\

F-statistic
&
--
&
--
&
0.2504
&
0.00180
&
0.0561
\\

Prob(F-statistic)
&
--
&
--
&
0.6168
&
0.966
&
0.813
\\

AIC
&
$1.854\times10^{5}$
&
--
&
--
&
--
&
--
\\

Durbin--Watson
&
1.954
&
--
&
--
&
--
&
--
\\

Log-Likelihood
&
--
&
--
&
$-1.106\times10^5$
&
$-1.1163\times10^5$
&
$-1.1141\times10^5$
\\

\bottomrule
\end{tabular}

\vspace{0.4em}

\footnotesize{
Values shown are regression coefficients. Confidence intervals are 95\% confidence intervals. The first two columns report primary ordinary least squares models using 7-day average temperature exposure and residualized temperature-shock exposure, respectively. The remaining columns report robustness analyses using fixed-effects, ZIP-aggregated, and two-way demeaned specifications. Fixed-effects models were estimated using repeated ZIP-level observations and within-location variation over time. ZIP-aggregated models used ZIP-level averages of cognitive difficulty and temperature exposure. Two-way demeaned models removed ZIP-level and month-level mean structure before estimation. Across all specifications, estimated temperature effects remained close to zero and confidence intervals included the null.
}

\end{table}

\textbf{ Extreme-weather analyses
}
Extreme-weather models were estimated to evaluate whether discrete severe-weather events showed stronger associations than continuous temperature-based exposure measures.

\clearpage
\section*{Supplementary Appendix C. Pathway and Heterogeneity Analyses}

\subsection*{Psychological pathway analyses for Attention}

\begin{longtable}{p{5cm}p{2.8cm}cccc}
\caption{Attention-related analyses}
\label{tab:attention_combined}
\\

\toprule
Analysis &
Variable / Component &
Estimate &
Std.Err &
95\% CI &
p-value \\
\midrule
\endfirsthead

\multicolumn{6}{c}{\textit{Table S3 continued}}\\
\toprule
Analysis &
Variable / Component &
Estimate &
Std.Err &
95\% CI &
p-value \\
\midrule
\endhead

\multicolumn{6}{l}{\textbf{Panel A. Main association analyses}}\\
\midrule

Baseline pooled model
& Intercept
& -0.7882
& 0.015
& (-0.817,-0.759)
& $<0.001$ \\

Baseline pooled model
& temp\_shock
& 0.0315
& 0.015
& (0.003,0.060)
& 0.032 \\

Baseline pooled model
& age
& -0.0232
& $<0.001$
& (-0.024,-0.023)
& $<0.001$ \\

\addlinespace

Lagged exposure model
& Intercept
& -0.6627
& 0.045
& (-0.751,-0.575)
& $<0.001$ \\

Lagged exposure model
& temp\_shock\_lag1
& 0.0498
& 0.045
& (-0.039,0.138)
& 0.270 \\

Lagged exposure model
& age
&  -0.0281
& 0.001
& (-0.030,-0.026)
& $<0.001$ \\

\addlinespace

Demographic-adjusted model
& temp\_shock
& 0.0328
& 0.0148
& (0.004,0.062)
& 0.027 \\

Quadratic specification
& temp\_shock$^{2}$
& -0.0201
& 0.0209
& (-0.061,0.021)
& 0.332 \\

\midrule
\multicolumn{6}{l}{\textbf{Panel B. Fixed-effects and location-based analyses}}\\
\midrule

ZIP3 fixed-effects model
& temp\_shock
& 0.011
& 0.021
& (-0.030,0.052)
& 0.617 \\

ZIP-by-month demeaned model
& temp\_shock
& 0.006
& 0.024
& (-0.041,0.053)
& 0.813 \\

Between-location averages
& temp\_shock
& 0.001
& 0.019
& (-0.036,0.038)
& 0.966 \\

Elderly subgroup fixed-effects model
& temp\_shock
& 0.003
& 0.041
& (-0.077,0.083)
& 0.974 \\

\midrule
\multicolumn{6}{l}{\textbf{Panel C. Heterogeneity analyses}}\\
\midrule

Interaction analysis
& Temp shock $\times$ drinking status
& -0.0063
& 0.0362
& (-0.077,0.065)
& 0.862 \\

Interaction analysis
& Temp shock $\times$ elderly status
& 0.0170
& 0.0403
& (-0.062,0.096)
& 0.673 \\

Subgroup analysis
& Elderly subgroup
& 0.0440
& 0.0367
& (-0.028,0.116)
& 0.234 \\

Subgroup analysis
& Non-elderly subgroup
& 0.0270
& 0.0158
& (-0.004,0.058)
& 0.092 \\

\midrule
\multicolumn{6}{l}{\textbf{Panel D. Mediation analyses}}\\
\midrule

Mental health pathway
& Temperature shock $\rightarrow$ mental health
& -0.0146
& --
& (-0.046,0.017)
& 0.361 \\

Mental health pathway
& Mental health $\rightarrow$ attention
& -0.2225
& --
& (-0.245,-0.200)
& $<0.001$ \\

Mental health pathway
& ACME
& 0.000382
& --
& (-0.000633,0.001023)
& 0.400 \\

Mental health pathway
& ADE
& 0.005449
& --
& (-0.003796,0.012448)
& 0.100 \\

\addlinespace

Fatigue pathway
& Temperature shock $\rightarrow$ fatigue
& -0.0076
& --
& (-0.088,0.073)
& 0.853 \\

Fatigue pathway
& Fatigue $\rightarrow$ attention
& 0.2040
& --
& (0.187,0.221)
& $<0.001$ \\

Fatigue pathway
& ACME
& -0.000427
& --
& (-0.001763,0.000984)
& 0.700 \\

Fatigue pathway
& ADE
& 0.007919
& --
& (-0.002195,0.015907)
& 0.200 \\

\addlinespace

Coping pathway
& Temperature shock $\rightarrow$ coping
& -0.0187
& --
& (-0.052,0.014)
& 0.267 \\

Coping pathway
& Coping $\rightarrow$ attention
& 0.4893
& --
& (0.460,0.518)
& $<0.001$ \\

Coping pathway
& ACME
& -0.001062
& --
& (-0.002601,0.000345)
& 0.100 \\

Coping pathway
& ADE
& 0.005359
& --
& (0.000887,0.012632)
& 0.001 \\

\addlinespace

Exploratory pathway
& PM$_{2.5}$ exposure
& -0.0186
& --
& (-0.022,-0.015)
& $<0.001$ \\

Exploratory pathway
& Blizzard exposure
& -0.0360
& --
& (-0.098,0.026)
& 0.252 \\

Exploratory pathway
& Biomarker mediation analyses
& Small estimates
& --
& Included 0
& n.s. \\

Exploratory pathway
& SBP / DBP / HbA1c mediation
& Small estimates
& --
& Included 0
& n.s. \\

\midrule
\multicolumn{6}{l}{\textbf{Panel E. Model diagnostics}}\\
\midrule

Baseline model
& Observations
& 370,307
& --
& --
& -- \\

Baseline model
& Log-likelihood
& -136,060
& --
& --
& -- \\

Baseline model
& LLR p-value
& --
& --
& --
& $<0.001$ \\

Lagged model
& Observations
& 40,000
& --
& --
& -- \\

Lagged model
& Log-likelihood
& -14,230
& --
& --
& -- \\

\bottomrule

\end{longtable}

\footnotesize{
Panel A summarizes primary and sensitivity analyses of temperature perturbations and attention outcomes. Panel B presents fixed-effects and location-based analyses designed to isolate within-location temporal variation. Panel C reports susceptibility and heterogeneity analyses. Panel D summarizes mediation analyses involving mental-health, fatigue-related, coping-related, environmental, and biomarker pathways. ACME denotes the average causal mediation effect and ADE denotes the average direct effect. Biomarker-related mediation analyses produced small indirect effects with confidence intervals spanning zero. ``n.s.'' indicates statistically non-significant associations ($p \geq 0.05$).
}

\subsection*{Psychological pathway analyses for Cognition}

\begin{longtable}{p{3cm}p{3.0cm}cccc}
\caption{Cognitive difficulty outcome analyses}
\label{tab:cognition_combined}
\\

\toprule
Analysis &
Variable / Component &
Estimate &
Std.Err &
95\% CI &
p-value \\
\midrule
\endfirsthead

\multicolumn{6}{c}{\textit{Table S4 continued}}\\
\toprule
Analysis &
Variable / Component &
Estimate &
Std.Err &
95\% CI &
p-value \\
\midrule
\endhead

\multicolumn{6}{l}{\textbf{Panel A. Psychological correlates}}\\
\midrule

Psychological correlate
&
Mental health burden
&
-0.114
&
0.004
&
--
&
$<0.001$
\\

Psychological correlate
&
Loneliness
&
0.053
&
--
&
--
&
$<0.001$
\\

Psychological correlate
&
Social functioning
&
-0.102
&
--
&
--
&
$<0.001$
\\

Psychological correlate
&
Literacy-related measure
&
-0.082
&
--
&
--
&
$<0.001$
\\

\midrule
\multicolumn{6}{l}{\textbf{Panel B. Environmental correlates}}\\
\midrule

Environmental correlate
&
Cumulative blizzard exposure
&
-0.0003
&
0.001
&
(-0.001,0.001)
&
0.652
\\

OLS environmental model
&
Intercept
&
0.1372
&
0.001
&
(0.134,0.140)
&
$<0.001$
\\

OLS environmental model
&
$temp\_anomaly$
&
0.0004
&
$4.67\times10^{-5}$
&
(0.000,0.001)
&
$7.17\times10^{-36}$
\\

OLS environmental model
&
$pm25$
&
-0.0016
&
$<0.001$
&
(-0.002,-0.001)
&
$<0.001$
\\

PM$_{2.5}$ mediation
&
$temp \rightarrow PM_{2.5} \rightarrow cognition$
&
$7.86\times10^{-8}$
&
--
&
--
&
Near zero
\\

PM$_{2.5}$ mediation
&
$blizzard \rightarrow PM_{2.5} \rightarrow cognition$
&
$\approx -3.14\times10^{-2}$
&
--
&
--
&
Near zero
\\

Environmental model
&
Observations
&
432,786
&
--
&
--
&
--
\\

Environmental model
&
Degrees of freedom (model)
&
2
&
--
&
--
&
--
\\

Environmental model
&
Degrees of freedom (residual)
&
432,783
&
--
&
--
&
--
\\

Environmental model
&
$R^2$
&
0.000
&
--
&
--
&
--
\\

Environmental model
&
Adjusted $R^2$
&
0.000
&
--
&
--
&
--
\\

Environmental model
&
F-statistic
&
80.94
&
--
&
--
&
--
\\

Environmental model
&
Prob(F-statistic)
&
$7.17\times10^{-36}$
&
--
&
--
&
--
\\

Environmental model
&
Log-Likelihood
&
$-1.3365\times10^5$
&
--
&
--
&
--
\\

Environmental model
&
AIC
&
$2.673\times10^5$
&
--
&
--
&
--
\\

Environmental model
&
Durbin-Watson
&
0.673
&
--
&
--
&
--
\\

\midrule
\multicolumn{6}{l}{\textbf{Panel C. Behavioral mediation analyses}}\\
\midrule

Behavior mediation
&
Smoking frequency
&
0.000753
&
--
&
--
&
Indirect effect
\\

Behavior mediation
&
Lifetime smoking exposure
&
0.000000
&
--
&
--
&
Indirect effect
\\

Behavior mediation
&
Alcohol use
&
0.000359
&
--
&
--
&
Indirect effect
\\

Behavior mediation
&
Smoking frequency ($a$ path)
&
0.0103
&
--
&
--
&
--
\\

Behavior mediation
&
Smoking frequency ($b$ path)
&
0.0730
&
--
&
--
&
--
\\

Behavior mediation
&
Lifetime smoking ($a$ path)
&
0.0000
&
--
&
--
&
--
\\

Behavior mediation
&
Lifetime smoking ($b$ path)
&
0.0000
&
--
&
--
&
--
\\

Behavior mediation
&
Alcohol use ($a$ path)
&
-0.0169
&
--
&
--
&
--
\\

Behavior mediation
&
Alcohol use ($b$ path)
&
-0.0213
&
--
&
--
&
--
\\

 \midrule
\multicolumn{6}{l}{\textbf{Panel D. Mental-health mediation analyses}}\\
\midrule

Mental-health mediation
&
Blizzard$_{30}$ $\rightarrow$ Mental health
&
0.0020
&
--
&
(-0.002,0.006)
&
0.336
\\

Mental-health mediation
&
Mental health $\rightarrow$ Cognitive difficulty
&
-0.1141
&
--
&
(-0.121,-0.107)
&
$<0.001$
\\

Mental-health mediation
&
Direct effect
&
-0.0003
&
--
&
(-0.001,0.001)
&
0.652
\\

Mental-health mediation
&
ACME
&
-0.000178
&
--
&
(-0.004564,0.003887)
&
0.924
\\

Mental-health mediation
&
ADE
&
-0.000250
&
--
&
(-0.001463,0.000766)
&
0.672
\\

Mental-health mediation
&
Total effect
&
-0.000428
&
--
&
(-0.004805,0.003632)
&
0.868
\\

Mental-health mediation
&
Proportion mediated
&
0.925
&
--
&
(-1.139,5.115)
&
0.128
\\

\midrule
\multicolumn{6}{l}{\textbf{Panel E. Sensitivity analyses}}\\
\midrule

Residualized temperature model
&
$temp\_7day_r$
&
$8.647\times10^{-6}$
&
--
&
(-0.001,0.001)
&
0.974
\\

Residualized blizzard model
&
$blizzard_r$
&
0.0008
&
--
&
(-0.006,0.008)
&
0.814
\\

Predicted mental-health model
&
Predicted mental health
&
-0.571
&
--
&
--
&
0.417
\\

Residualized temperature model
&
$R^2$
&
0.000
&
--
&
--
&
--
\\

Residualized temperature model
&
Adjusted $R^2$
&
-0.000
&
--
&
--
&
--
\\

Residualized temperature model
&
F-statistic
&
0.00104
&
--
&
--
&
--
\\

Residualized temperature model
&
Prob(F-statistic)
&
0.974
&
--
&
--
&
--
\\

Residualized temperature model
&
Log-Likelihood
&
607.66
&
--
&
--
&
--
\\

Residualized blizzard model
&
$R^2$
&
0.000
&
--
&
--
&
--
\\

Residualized blizzard model
&
Adjusted $R^2$
&
-0.000
&
--
&
--
&
--
\\

Residualized blizzard model
&
F-statistic
&
0.055
&
--
&
--
&
--
\\

Residualized blizzard model
&
Prob(F-statistic)
&
0.814
&
--
&
--
&
--
\\

\bottomrule

\end{longtable}

\footnotesize{
Panel A summarizes psychological and psychosocial correlates of subjective cognitive difficulty. Panel B reports environmental correlates, PM$_{2.5}$ pathway analyses, and environmental-model diagnostics. Panel C summarizes behavioral mediation analyses involving smoking and alcohol-related variables. Panel D reports mental-health mediation analyses linking cumulative blizzard exposure and cognitive difficulty. Panel E presents sensitivity analyses using residualized exposure measures and predicted mental-health scores.
}

 \clearpage

\begin{figure}[htbp]
\centering

\begin{tikzpicture}[
node distance=3cm,
box/.style={
draw,
rounded corners,
minimum width=2.8cm,
minimum height=0.8cm,
align=center
},
arrow/.style={->, thick}
]

\node[box] (lonely)
{Loneliness};

\node[box,right=of lonely]
(mh)
{Mental Health};

\node[box,right=of mh]
(cog)
{Subjective\ Cognitive Difficulty};

\draw[arrow] (lonely) -- (mh);
\draw[arrow] (mh) -- (cog);
\draw[arrow,bend left=20] (lonely) to (cog);

\end{tikzpicture}

\caption{
Conceptual psychosocial pathway evaluated in BRFSS replication analyses. The direct and indirect pathways represent the decomposition examined in mediation analyses.
}

\label{fig:brfss_pathway}

\end{figure}

 \section*{Supplementary Appendix D. Exploratory Machine Learning and Latent Representation Analyses}

These analyses were conducted as exploratory extensions to the primary longitudinal analyses. To examine whether latent psychological structure was associated with attention and subjective cognitive difficulty under short-term environmental stress exposure, we implemented an exploratory framework combining latent representation learning, mediation analyses, heterogeneous treatment effect estimation, interaction modeling, and semantic embedding approaches. Because loneliness, social health, and mental health measures were moderately correlated, a conditional variational autoencoder (cVAE) was used to learn lower-dimensional latent psychological representations:

\[
(T,X,M)
\rightarrow
(Z_1,Z_2)
\]

where \(T\) denotes temperature shock exposure, \(X\) denotes observed covariates, \(M\) denotes observed psychological measures, and \((Z_1,Z_2)\) denote learned latent representations.

\paragraph{Semantic embedding analyses}

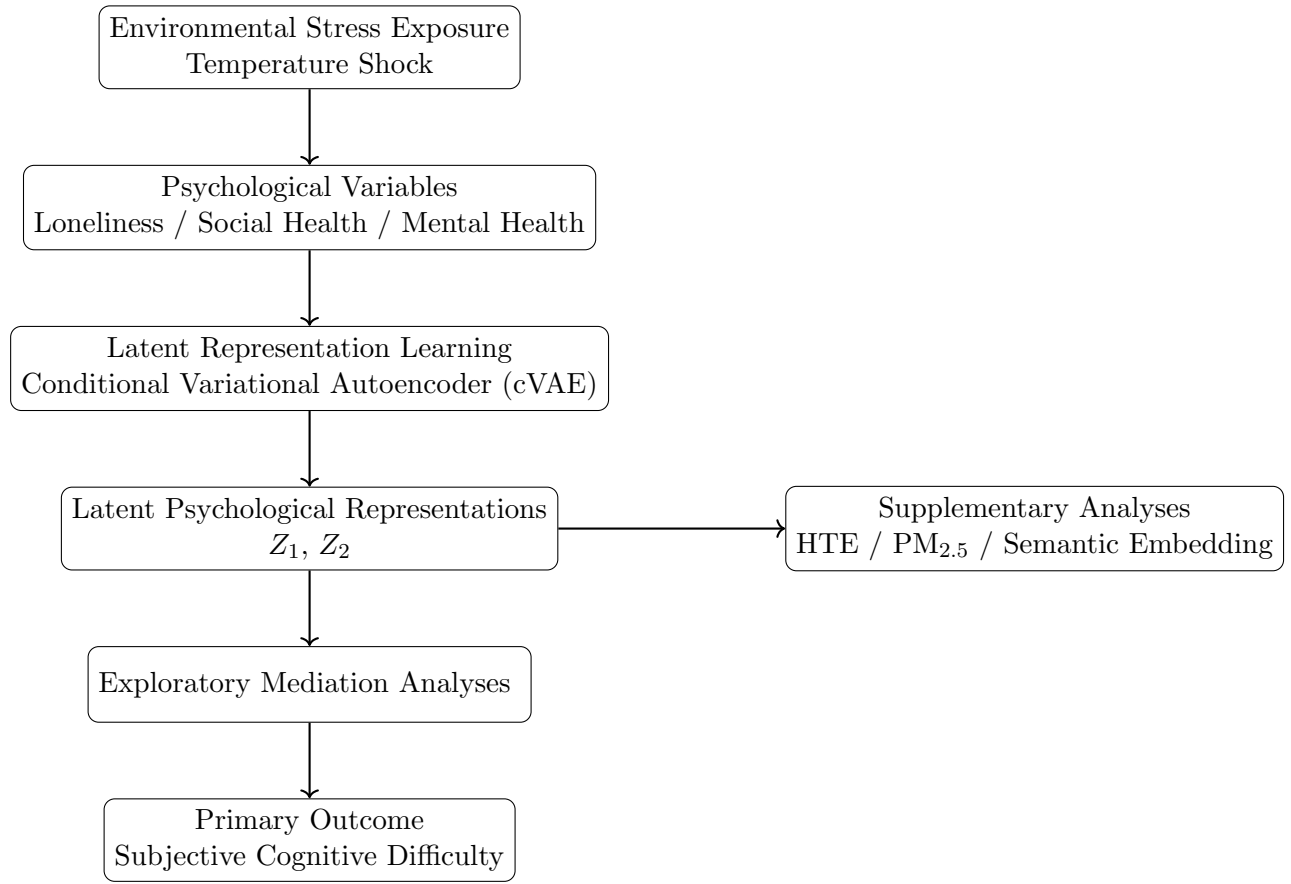
\begin{figure}[H]
\centering

\begin{tikzpicture}[
node distance=1cm,
box/.style={
draw,
rounded corners,
minimum width=2cm,
minimum height=1cm,
align=center,
font=\small
},
arrow/.style={->, thick}
]

\node[box] (exp) {
Environmental Stress Exposure\\
Temperature Shock
};

\node[box, below=of exp] (psych) {
Psychological Variables\\
Loneliness / Social Health / Mental Health
};

\node[box, below=of psych] (cvae) {
Latent Representation Learning\\
Conditional Variational Autoencoder (cVAE)
};

\node[box, below=of cvae] (latent) {
Latent Psychological Representations\\
$Z_1$, $Z_2$
};

\node[box, below=of latent] (med) {
Exploratory Mediation Analyses
};

\node[box, below=of med] (outcome) {
Primary Outcome\\
Subjective Cognitive Difficulty
};

\node[
box,
right=3cm of latent,
minimum width=3cm
] (supp) {
Supplementary Analyses\\
HTE / PM$_{2.5}$ / Semantic Embedding
};

\draw[arrow] (exp) -- (psych);
\draw[arrow] (psych) -- (cvae);
\draw[arrow] (cvae) -- (latent);
\draw[arrow] (latent) -- (med);
\draw[arrow] (med) -- (outcome);
\draw[arrow] (latent.east) -- (supp.west);

\end{tikzpicture}

\caption{
Psychological and psychosocial variables showed substantially larger associations with subjective cognitive difficulty than environmental exposure measures.}

\label{fig:cognition_framework}

\end{figure}

\begin{table}[p]
\centering
\caption{Exploratory machine-learning, latent mediation, and replication analyses}
\label{tab:exploratory_ml_combined}
\scriptsize

\begin{tabular}{lcc}
\toprule
\multicolumn{3}{l}{\textbf{Panel A. Exploratory latent mediation analyses}}\\
\midrule

Quantity
&
Attention
&
Subjective cognitive difficulty
\\

\midrule

Indirect effect (NIE)
&
-0.00053
&
-0.00181
\\

$a_1$
&
--
&
0.00597
\\

$a_2$
&
--
&
-0.00908
\\

$b_1$
&
--
&
-0.31150
\\

$b_2$
&
--
&
-0.00537
\\

\bottomrule
\end{tabular}

\vspace{0.6em}


\begin{tabular}{lcc}
\toprule
\multicolumn{3}{l}{\textbf{Panel B. Additional exploratory machine-learning analyses}}\\
\midrule

Analysis
&
Quantity
&
Estimate
\\

\midrule

CausalForestDML
&
Mean treatment effect
&
$-4.89\times10^{-6}$
\\

CausalForestDML
&
Standard deviation
&
$5.06\times10^{-4}$
\\

PM$_{2.5}$ interaction model
&
Interaction coefficient ($\beta_3$)
&
$-1.61\times10^{-5}$
\\

\bottomrule
\end{tabular}

\vspace{0.6em}


\begin{tabular}{lccc}
\toprule
\multicolumn{4}{l}{\textbf{Panel C. Replication analyses in BRFSS}}\\
\midrule

Analysis &
Estimate &
95\% CI Lower &
95\% CI Upper \\
\midrule

Loneliness $\rightarrow$ Mental-health burden
&
2.676
&
--
&
--
\\

Loneliness $\rightarrow$ SCD
&
OR=1.81
&
--
&
--
\\

Mental-health burden $\rightarrow$ SCD
&
OR=1.08/day
&
--
&
--
\\

ACME
&
0.011
&
0.010
&
0.012
\\

ADE
&
0.027
&
0.026
&
0.028
\\

Total effect
&
0.038
&
0.038
&
0.039
\\

Proportion mediated
&
28.6\%
&
26.9\%
&
30.1\%
\\

\bottomrule
\end{tabular}

\vspace{0.4em}

\footnotesize{
Panel A reports exploratory latent mediation analyses using latent psychological representations learned through conditional variational autoencoders (cVAEs). Panel B reports additional exploratory machine-learning analyses, including heterogeneous treatment-effect estimation using CausalForestDML and PM$_{2.5}$ interaction models. Panel C reports replication analyses conducted in BRFSS, including mediation estimates and associations among loneliness, mental-health burden, and subjective cognitive decline (SCD).
}

\end{table}

\end{document}